\documentclass[12pt]{article}
\usepackage{epsfig}
\begin{document}

\def\calO{{\cal O}}
\def\lsim{\raisebox{-1mm}{$\stackrel{<}{\sim}$}}

\begin{titlepage}

\begin{center}

\vspace{5mm}

\Large{\bf Two-pion-exchange and other higher-order contributions to \\
the $pp\to pp\pi^0$ reaction }

\vspace{1cm}
  {\large Y. Kim$^{(a,b)}$,
  T. Sato$^{(c)}$, F.  Myhrer$^{(a)}$ and
  K. Kubodera$^{(a)}$}

\vskip 0.5cm
(a)~{\large Department of Physics and Astronomy,
University of South Carolina, Columbia,
South Carolina 29208, USA}

(b) {\large School of Physics,
Korea Institute for Advanced Study,
Seoul 130-012, Korea}

(c)   {\large  Department of Physics, Osaka University,
Toyonaka, Osaka 560-0043, Japan}

\end{center}

\centerline{(\today) }

\vskip 1cm

\begin{abstract} 
Much effort has been invested 
on effective-field-theoretical studies of 
the near-threshold $NN\!\!\to\!\!NN\pi$ reactions
and, in order to deal with the somewhat 
large three-momentum transfers involved, 
the momentum counting scheme (MCS) was 
proposed as an alternative to the usual Weinberg counting scheme. 
Given the fact that a quantitative explanation
of the existing high-precision $NN\!\!\to\!\! NN\pi$ data
requires a careful examination of higher chiral order contributions 
to the transition operator,
we make a detailed numerical investigation of the convergence property
of MCS for a pilot case of the $pp\to pp\pi^0$ reaction.
Our study indicates that MCS is superior to the Weinberg scheme 
in identifying dominant higher order contributions
to the $NN\!\! \to\!\! NN\pi$ reactions.

\end{abstract}

\end{titlepage}

\newpage

There exists a substantial accumulation of high-precision data
on various observables for the near-threshold $NN\!\!\to\!\!NN\pi$
reactions, e.g. \cite{meyeretal90},
and providing a coherent understanding of these experimental results 
has been a prominent theoretical challenge, see e.g., 
\cite{ms91,lr93,pmmmk96,cfmv96,am01,hk02,hanhart04, Lensky05}. 
Heavy-baryon chiral perturbation theory (HB$\chi$PT), 
which is a low-energy effective field theory of QCD,  
is expected to offer a systematic framework 
for describing these reactions. 
In HB$\chi$PT,  the four-momentum $Q$ characterizing 
a given physical process is assumed to be small 
compared to the chiral scale 
$\Lambda_\chi \simeq 4\pi f_\pi \simeq 1$ GeV,
and contributions to the transition amplitude are classified according to 
the power (chiral order) of the expansion parameter 
$\epsilon \equiv Q/\Lambda_\chi$.\footnote{ 
The numerical value of the chiral expansion parameter
is $\epsilon \simeq m_\pi / \Lambda_\chi \simeq 1/7.1$,
while the ``recoil-correction" expansion parameter is 
$m_\pi / m_N \simeq 1/6.7$.  
Since these parameters have roughly the same numerical value, 
the chiral and recoil-correction expansions are 
combined together in HB$\chi$PT. } 
The coefficients of possible terms in the HB$\chi$PT
Lagrangian, called low-energy constants (LECs),
can in principle be linked to 
the matrix elements of QCD operators but in practice they
are determined from experimental observables. 
Once all the LEC are fixed up to a specified chiral order,
HB$\chi$PT allows us to make predictions 
for a wide range of hadronic and electroweak processes. 

Although, as mentioned, HB$\chi$PT presupposes the smallness of 
its expansion parameter $Q/\Lambda_\chi$,
the pion production reactions involve somewhat large 
momentum transfers, $p\!\simeq\!\sqrt{m_\pi m_N}$,
even at threshold.  This implies that
the application of HB$\chi$PT to 
the $NN\!\!\to\!\!NN\pi$ reactions
may involve some delicate aspects, but
this also means that these processes may serve as a good 
test case for probing the applicability 
(or the limit of applicability) of HB$\chi$PT.
It is worth emphasizing that finding 
a valid EFT expansion scheme 
for the $NN\!\!\to\!\!NN\pi$ reactions 
is of general importance (going beyond the specific context 
of the $NN\!\!\to\!\!NN\pi$ reactions)
because, once such a scheme is found, 
we expect to be able to develop similar frameworks
for other nuclear processes which involve 
rather large energy-momentum transfers
and hence may not be quite amenable to the straightforward application 
of the ordinary HB$\chi$PT approach.
To account for the rather large momentum transfers
involved in the $NN\!\!\to\!\!NN\pi$ reactions,
Cohen, Friar, Miller and van Kolck ~\cite{cfmv96} 
proposed to replace the ordinary chiral counting scheme of Weinberg 
(W-scheme) with a new scheme called
the momentum counting scheme (MCS). 
The expansion parameter in MCS is 
$\tilde{\epsilon}\equiv p/m_N\!\simeq\!(m_\pi / m_N)^{1/2}
\simeq 1/2.6$,
which is appreciably larger
than the expansion parameter
$\epsilon \simeq m_\pi / m_N\simeq 1/6.7$ in the W-scheme.

Although the formal aspects of MCS have been discussed
rather extensively~\cite{hanhart04},
it seems fair to say that its practical utility is yet to be established.
An important point to be noted here is that, whereas in the W-scheme
a given Feynman diagram corresponds to a definite power
in $\epsilon$ (it thus has a unique HB$\chi$PT order),
this type of correspondence does not in general exist in MCS, 
because a given Feynman diagram 
can involve contributions that belong to different orders in 
$\tilde{\epsilon}$.
In the following, the contribution 
corresponding to the lowest power of $\tilde{\epsilon}$
for a given diagram is referred to as 
the ``leading term"~\cite{hk02,hanhart04},
and the remaining (higher order in $\tilde{\epsilon}$) 
contributions of the diagram 
as ``sub-leading terms".
It is to be emphasized
that, in order to study the convergence property of MCS,
we need to examine not only the behavior of the leading terms
but also that of the sub-leading terms.
Our first attempt at such a study 
was described in Ref.~\cite{fm06,ksmk07}, to be referred to as KSMK1.
The results presented in KSMK1
indicate that the individual terms 
in the MCS expansion exhibit much more complicated behavior
than the straightforward power counting 
in $\tilde{\epsilon}$ would indicate.

The study in KSMK1, however, is subject to elaboration
in at least two points.  
The first is that, for the nucleon propagator,
KSMK1 used the standard HB$\chi$PT propagator
instead of the MCS propagator (see below), 
and the consequences of this eclectic treatment
need to be investigated.
The second point is that, in KSMK1, 
only the nucleons and pions appear as effective 
degrees of freedom, with the $\Delta$-particle field integrated away. 
In view of the rather high momentum transfers involved 
in the $NN\!\!\to\!\!NN\pi$ reactions, 
the influence of including the $\Delta$-particle
as an explicit degree of freedom is well worth examination.
Since however this second point has been discussed in
Refs.~\cite{hanhart04,Baruetal08},\footnote{
In Ref.~\cite{hanhart04} it is convincingly 
argued that at next-to-leading-order (NLO) 
in MCS, the diagrams involving a $\Delta$ cancel
for s-wave pion production, i.e. 
$\Delta$ will only enter at higher orders in MCS.} 
we concentrate here on the first point.
Furthermore, we limit ourselves here
to the study of the transition operators 
(or equivalently the transition amplitudes in plane-wave
approximation) even though
the actual transition amplitudes need to be calculated
with the use of the distorted-waves (DW) for the initial and final
two-nucleon systems. 
Previous studies (see, e.g.,~\cite{pmmmk96,cfmv96,hanhart04,slmk})
have shown that, in order to reproduce the experimental data,
it is crucially important to evaluate higher chiral-order 
contributions to the transition operator. 
In the present work, therefore,
we focus on the convergence property of the MCS expansion
of the transition amplitudes calculated in plane-wave approximation.
As mentioned, 
the $NN\!\!\to\!\!NN\pi$ reactions serve as a pilot case 
for nuclear reactions that involve 
rather high energy-momentum transfers,
and it is hoped that our present study
(despite its stated limitations) will shed some light 
on the general issue of an effective-field theoretical 
treatment of those reactions. 

Of the various possible isospin channels  for
the $NN\!\!\to\!\!NN\pi$ reactions,
we concentrate (as we did in KSMK1)
on the $pp\!\!\to\!\! pp\pi^0$ reaction.
The amplitudes for the $NN\!\!\to\!\!NN\pi$ reactions
are in general dominated by 
a pion-rescattering diagram 
in which rescattering
is caused by the Weinberg-Tomozawa (WT) interaction term,
but this particular diagram 
does not contribute to the 
$pp\!\to\!pp\pi^0$ reaction.
This feature makes the $pp\!\to\!pp\pi^0$ reaction
uniquely suited for investigating the behavior
of higher order terms
in $s$-wave pion production.\footnote{
It should be mentioned, however,  
that the $s$-wave pion production amplitude
represents a more complicated case than 
the $p$-wave pion production amplitude;
the convergence property of MCS for the latter
has been discussed in Ref.\cite{hvm00}.}
%

\begin{figure}
\begin{center} 
\includegraphics[width=3cm]{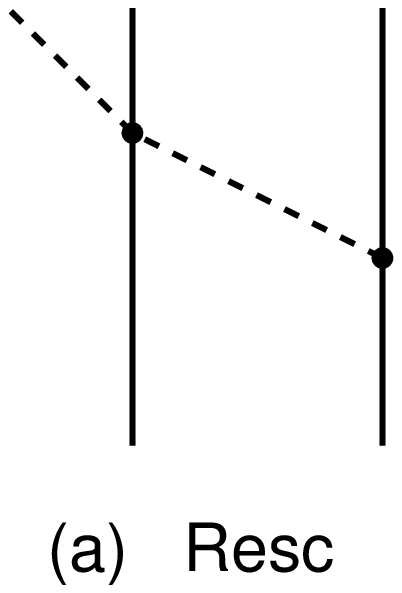}\hspace*{1cm}
\includegraphics[width=3cm]{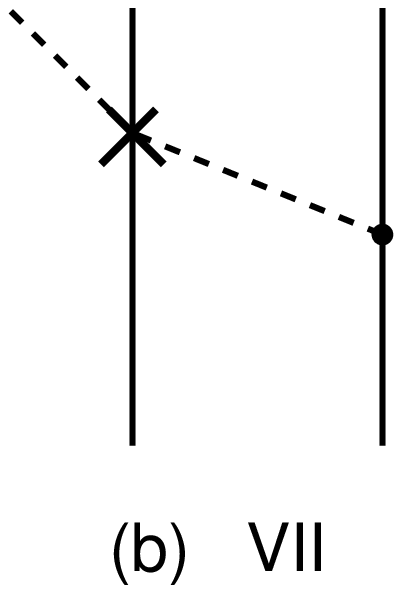}\hspace*{1cm} 
\caption[]{ \protect 
One-pion-exchange rescattering diagrams
of order $\tilde{\epsilon}^3$ (in MCS) for the
threshold $pp\!\!\to\!\! pp\pi^0$ reaction.
Diagram 1(a) represents
the lowest-order non-vanishing
one-pion-exchange contribution (NLO in the W-scheme) ,
while diagram 1(b) represents its recoil correction 
(NNLO in the W-scheme).
Adopting the same convention as in Ref.~\cite{dkms99}
for labeling the diagrams, 
we call diagram 1(a) the {\it rescattering} diagram, 
or the {\it Resc} diagram for short,
and refer to diagram 1(b) as diagram VII.$^3$ 
The re-scattering vertex in the 
{\it Resc} diagram comes from 
${\cal L}^{(2)}_{\pi N} = 
c_1N^\dagger({\rm Tr}\chi_+)N + \cdots$~\cite{bkm95}, whereas 
diagram VII contains the re-scattering (recoil correction) 
vertex of the next chiral order lagrangian ${\cal L}^{(3)}_{\pi N}$. 
}
\end{center} 
\end{figure}
\begin{figure}
\begin{center} 
\includegraphics[width=3cm]{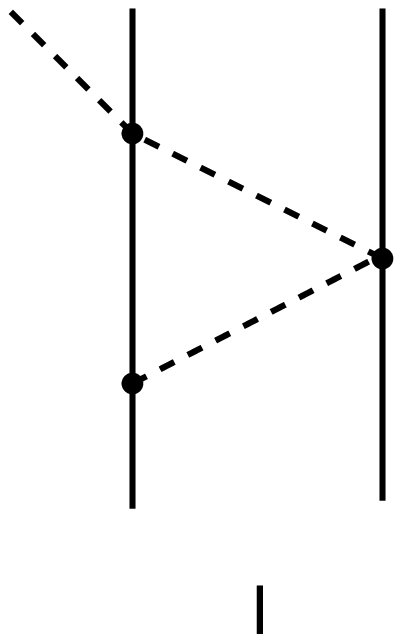}\hspace*{1cm}
\includegraphics[width=3cm]{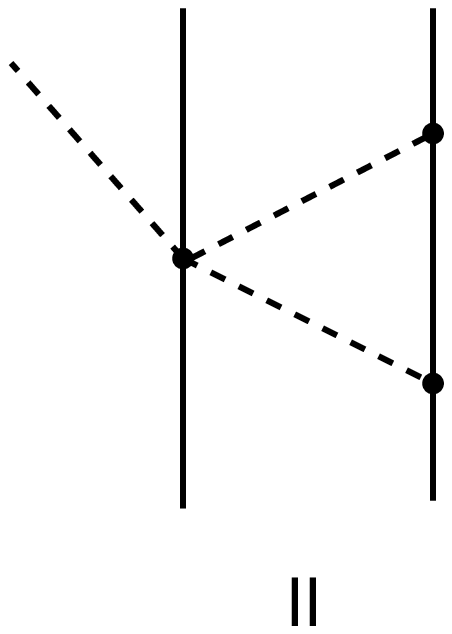}\hspace*{1cm}
\includegraphics[width=3cm]{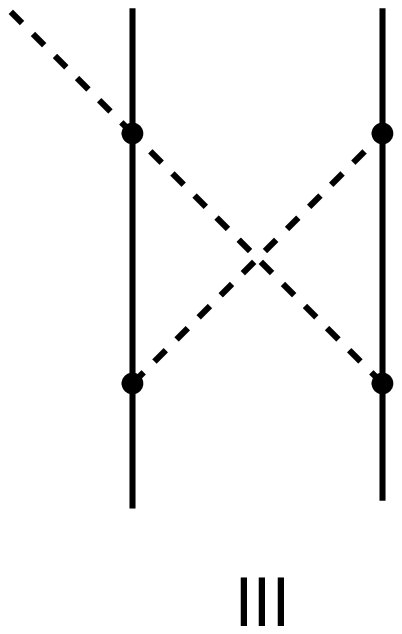}\hspace*{1cm}
\includegraphics[width=3cm]{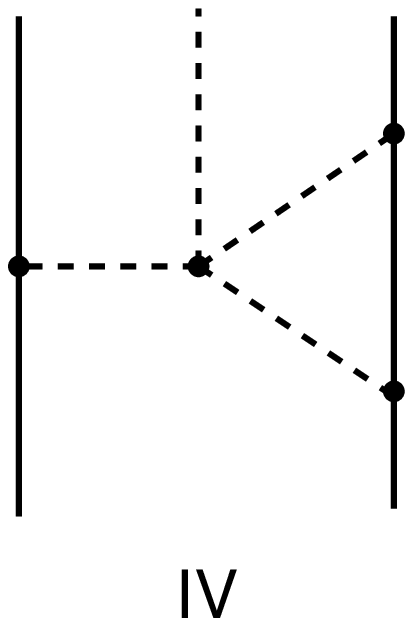} 
\caption[]{ \protect
The two-pion-exchange (TPE) diagrams
considered to be of order $\tilde{\epsilon}^2$ in 
the formal counting in MCS (NNLO in W-scheme). 
We only show one representative diagram
for each group, suppressing similar diagrams belonging 
to the same group (see Ref.~\cite{dkms99} for details). 
The diagram labels, I, II, III and IV,
follow the convention used in Ref.~\cite{dkms99}.
All the vertices arise from  the lowest chiral order lagrangians, 
${\cal L}^{(1)}_{\pi N}$ and ${\cal L}^{(2)}_{\pi \pi}$,
see Ref.~\cite{bkm95}.
}
\end{center}
\end{figure}
\begin{figure}
\begin{center}
\includegraphics[width=3cm]{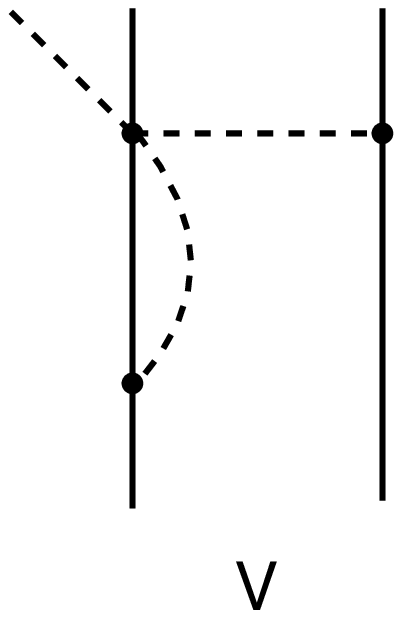}\hspace*{1cm}
\includegraphics[width=3cm]{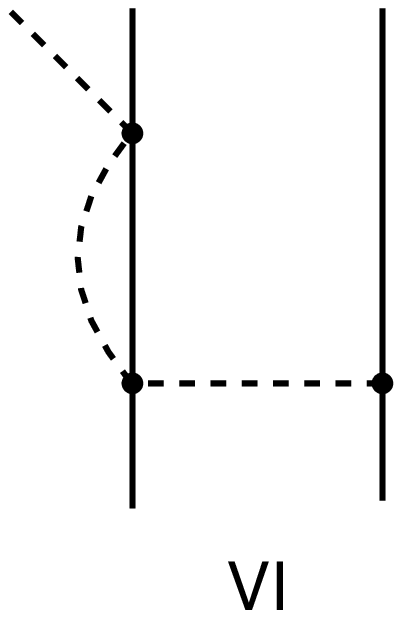}
\caption[]{ \protect 
One-pion-exchange diagrams with vertex correction 
belonging to order $\tilde{\epsilon}^5$ in MCS 
(NNLO in the W-scheme). 
The diagram labels, V and VI,
follow the convention used in Ref.~\cite{dkms99}.
}
\end{center}
\end{figure} 
\begin{figure}
\begin{center}
\includegraphics[width=3cm]{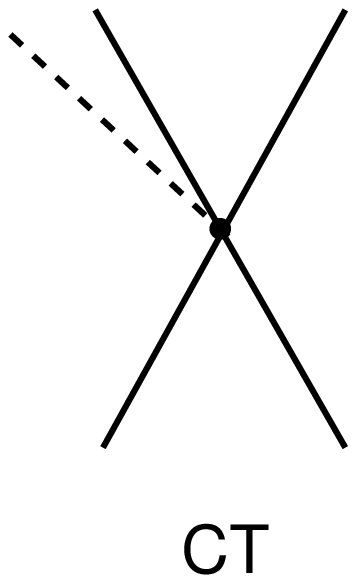}
\caption[]{\protect 
The counter-term (CT) diagram of order $\tilde{\epsilon}^3$ 
(NNLO in the W-scheme). 
}
\end{center}
\end{figure}

We have derived in Ref.~\cite{dkms99}
all the possible two-body transition operators
for $s$-wave pion production 
in the $pp\!\!\to\!\!pp\pi^0$ reactions up to 
next-to-next-to-leading order (NNLO) 
in the W-scheme.\footnote{The structure of the one-body 
transition operators is well known in HB$\chi$PT.}
The diagrams generating these operators
can be categorized into several groups:
the pion-rescattering diagram illustrated in Fig.~1(a)
and its recoil correction diagram (diagram VII) 
shown in Fig.~1(b);
the two-pion exchange (TPE) diagrams 
(diagrams I$\sim$IV) in  Fig.~2,
the vertex loop-correction diagrams
(diagrams V and VI) in Fig.~3, and
the contact-interaction diagram (CT diagram)
in Fig.~4.\footnote{In labeling the diagrams in 
Figs.~1$\sim$4,
we are following the same convention
as in Ref.~\cite{dkms99};
this leads to a somewhat awkward situation here
that the diagrams are not numbered 
according to the order of their appearance
in the text; 
``diagram VII" shows up earlier than diagrams 
I, II, III, etc.}
In what follows, the diagram in Fig.1(a), which specifically represents
the lowest-order non-vanishing
one-pion-exchange contribution to $pp\!\to\!pp\pi^0$,
shall be simply referred to 
as the 
{\it rescattering} diagram, or the {\it Resc} 
diagram for short. 
\begin{table}
\caption{\label{diagrams}\protect
The orders of diagrams in the W-scheme and MCS.
The labels of the diagrams conform with those used in 
Ref.~\cite{dkms99}. 
When a diagram is assigned 
${\cal O}(\tilde{\epsilon}^n)$ in MCS, it means that its ``leading term" is
${\cal O}(\tilde{\epsilon}^n)$ with the understanding
that there can also be terms higher order in $\tilde{\epsilon}$
(``sub-leading terms"). 
}
\vspace{6mm}
\begin{tabular}{|l || c|c|c | c|c|c|c|c|c|}
\hline 
Diagram & Resc & I & II&III&IV&V&VI&VII&CT\\
\hline\hline 
W-scheme &  NLO &  NNLO & NNLO & NNLO
&  NNLO & NNLO& NNLO& NNLO &  NNLO \\
\hline 
 MCS & ${\cal O}(\tilde{\epsilon}^3)$ & ${\cal O}(\tilde{\epsilon}^2)$ 
& ${\cal O}(\tilde{\epsilon}^2)$ & ${\cal O}(\tilde{\epsilon}^2)$ 
& ${\cal O}(\tilde{\epsilon}^2)$ & ${\cal O}(\tilde{\epsilon}^5)$ 
& ${\cal O}(\tilde{\epsilon}^5)$ & ${\cal O}(\tilde{\epsilon}^3)$ 
& ${\cal O}(\tilde{\epsilon}^3)$ \\
\hline
\end{tabular}
\vspace{5mm}
\end{table}
Table~\ref{diagrams} shows the power counting of these diagrams 
(Figs. 1 $\sim$ 4) in the W-scheme 
(powers in $\epsilon$) and in MCS 
(powers in $\tilde{\epsilon}$).
The table indicates that the two schemes give significantly 
different classifications of these diagrams.
Diagrams I $\sim$ VII and the CT diagram,
which are all categorized as NNLO terms in the W-scheme,
belong to different orders in MCS, 
ranging from ${\cal O}(\tilde{\epsilon}^2)$ 
to ${\cal O}(\tilde{\epsilon}^5)$.
Furthermore, the Resc term, which 
in the W-scheme is of NLO 
and thus of lower chiral order than any other terms 
in the table, belongs to ${\cal O}(\tilde{\epsilon}^3)$ in MCS, 
and thus of higher order in $\tilde{\epsilon}$
than Diagrams I $\sim$ IV.
These features invite us to investigate the actual numerical behavior 
of the diagrams listed in Table~\ref{diagrams} in the context of MCS
and examine whether MCS indeed provides 
a useful guide in organizing higher order contributions.
The main purpose of this article is to report on
such a numerical investigation.

\vspace{2mm}
A major difference between the W-scheme~\cite{hanhart04}
and MCS~\cite{hanhart04}
is associated with the different treatments of the
nucleon propagator (see e.g. Ref.~\cite{bkm95}),
and we briefly recapitulate this point here.
In HB$\chi$PT the nucleon momentum $Q^\mu$
is written as $Q^\mu \!=\!m_N \, v^\mu\!+\!q^\mu$, 
where $m_N$ is the nucleon mass, 
and $v^\mu$ is the four-velocity,
which may be chosen to be $v^\mu = (1;\vec{0})$;
it is assumed that $|q^\mu| \!\ll\! m_N$. 
In terms of $q_\mu$ the Feynman propagator 
for the heavy nucleon is expressed as 
\begin{eqnarray}
S_N(q) = i\; \frac{ \not\!\! Q +m_N }{Q^2-m_N^2} = i\; \left(
v\cdot q +\frac{q^2}{2m_N}\right)^{-1} \left( 
\frac{1+\gamma_0 }{2} +\frac{\gamma_0\; q_0 }{2m_N} 
- \frac{ \vec{\gamma}\cdot \vec{q} }{2m_N} 
\right) \; . 
\label{eq:Nprop}
\end{eqnarray}
In one version of non-relativistic HB$\chi$PT,
the free heavy-nucleon Lagrangian is chosen 
in such a manner that the heavy-nucleon propagator 
is given by
\begin{eqnarray}
S_N (q) = \frac{i}{v\cdot q } \; ,
\label{eq:Nprop1}
\end{eqnarray} 
and the difference between  
the propagators in
eqs.(\ref{eq:Nprop}) and (\ref{eq:Nprop1})
is treated as perturbative recoil corrections 
which  are accounted for in 
higher chiral order Lagrangians. 
For convenience, we refer to this approach 
as the W-scheme.\footnote{It should be remarked, however,
that Weinberg~\cite{weinberg91} discussed problems
associated with using the static nucleon propagator, 
Eq. (\ref{eq:Nprop1}),
in the two-nucleon systems, 
and that he proposed a possible remedy, which however upset
the original counting scheme.}
The work in Refs.~\cite{ksmk07,dkms99}
is based on the use of $S_N (q)$ given in eq.(\ref{eq:Nprop1}). 

%

Meanwhile, in MCS, one rearranges 
the expression in eq.(\ref{eq:Nprop}) to adapt it
to a specific kinematic situation pertaining to  
the $NN\!\!\to\!\!NN\pi$ reaction at threshold.
In order for a threshold pion to be produced
from two nucleons, 
the four-momenta of the incoming on-shell nucleons 1 and 2
(in the CM system) with 
$m_{\mbox{\tiny{$N$}}}v^\mu$ subtracted must be
$p_1\!=\!(m_\pi/2, \vec{p})$ 
and $p_2\!=\!(m_\pi/2, -\vec{p})$,
where
$|\vec{p}| \!=\! \sqrt{m_{\mbox{\tiny{$N$}}}m_\pi \!+\!m_\pi^2/4} 
\!=\!\sqrt{m_{\mbox{\tiny{$N$}}}m_\pi}[1\!+\tilde{\epsilon}^2\!/8+
{\cal O}(\tilde{\epsilon}^4)]$.
(Recall $\tilde{\epsilon}
\!\equiv\!\sqrt{m_\pi/{m_{\mbox{\tiny{$N$}}}}}$.)
In what follows, for notational simplicity,
we write $p$ instead of $p_1$.
When an incoming on-shell nucleon of four-momentum $p$ 
produces a pion of four-momentum $l$, 
the off-shell nucleon propagator has a momentum $p\!-\! l$,
where the loop-integral four-momentum $l$ 
behaves like:
$ v\cdot l = l_0\, 
\sim\,|\vec{p}|\! \simeq \!\!\sqrt{m_Nm_\pi}$,
$|\vec{l}|\,
\sim\,|\vec{p}| \simeq \!\!\sqrt{m_Nm_\pi}$.
Let the nucleon propagator in eq.(\ref{eq:Nprop}) 
be rewritten as
\begin{eqnarray}
S_N\!\!\!& &\!\!\!\!\!\!\!\!\!\!\!\!(p\!-\!l)\! =\! i \!\left( v\!\cdot\!(p-l) \!-\! 
\frac{(\vec{p} \!- \!\vec{l})^2 }{2m_N} 
\!+\!\frac{[v\!\cdot\!(p-l)]^2}{2m_N} \right)^{\!\!\!-1} \!\!
\left( 
\frac{1\!+\!\gamma_0 }{2}   
\!- \!\frac{ \vec{\gamma}\!\cdot\!(\vec{p} -\vec{l}) }{2m_N} 
\!+\!\frac{\gamma_0(p_0 -l_0) }{2m_N}
\right) 
\nonumber \\ 
\!\!\!\!\!\!\!\!\!\!\!&= &\!\!\!
i \!\left( \!- v\!\cdot\!l   \!+\!v\!\cdot\!p 
\!-\! \frac{\vec{p}^{\; 2} }{2m_N} 
\!-\! \frac{\vec{l}^{\; 2} \!-\!2\vec{l}\!\cdot\!\vec{p}}{2m_N} 
\!+\!\!\frac{ [p_0-l_0]^2 }{2m_N}
\right)^{\!\!\!-1} \!\!\!
\left( 
\frac{1\!+\!\gamma_0 }{2}   
\!-\! \frac{ \vec{\gamma}\!\cdot\!(\vec{p} -\vec{l}) }{2m_N} 
\!+\!\frac{\gamma_0 (p_0 -l_0) }{2m_N}
\right) 
\nonumber \\ && 
\label{eq:Nprop3} 
\end{eqnarray} 
Since $v\!\cdot\!p=\frac{m_\pi}{2}$ and
$v\!\cdot\!l\!\sim\!\sqrt{m_Nm_\pi}$, 
the difference 
\begin{eqnarray}
v\!\cdot\!p - \!\frac{\vec{p}^{\; 2} }{2m_N} 
=\frac{m_\pi}{2}
-\frac{m_Nm_\pi(1\!+\!\tilde{\epsilon}^2/4)}{2m_N}
=- \sqrt{m_N m_\pi}\,\,\frac{ \tilde{\epsilon}^3 }{8}\,,
\label{eq:cancellation}
\end{eqnarray} 
appearing in eq.(\ref{eq:Nprop3}) 
is of higher order in $\tilde{\epsilon}$ compared to $v\cdot l$ 
and the other terms in eq.(\ref{eq:Nprop3}) 
and therefore can be dropped. 
This treatment of the $\vec{p}^{\; 2}/2m_N$ term
is to be contrasted to that in the W-scheme,
where the $\vec{p}^{\; 2}/2m_N$ term is included in 
the higher chiral order lagrangian and 
treated as a perturbative correction. 
We now reorganize eq.(\ref{eq:Nprop3}) as 
\begin{eqnarray}
S_N (p\!-\!l)\!\!\!\!&=&\!\!\!\!
i \!\left[-v\!\cdot\!l\!-\! \left\{ 
 \;\frac{\!\vec{l}^{\; 2} \!-\!2\vec{l}\!\cdot\!\vec{p} }{2m_N} 
-\!\frac{( p_0 - l_0 )^2}{2m_N}\right\} 
\right]^{\!\!-1}\!\!\!
\left( 
\frac{1\!+\!\gamma_0 }{2}   
\!-\!\frac{ \vec{\gamma}\!\cdot\!(\vec{p} -\vec{l}) }{2m_N} 
\!+\!\frac{\gamma_0 (p_0 -l_0) }{2m_N}
\right) \!+ \cdots  
\nonumber \\ && 
\label{eq:Nprop4} 
\end{eqnarray} 
Here, compared with the leading term $v\! \cdot l$, 
the terms in the curly brackets are of order 
$\tilde{\epsilon}$ or higher. 
Expanding eq.(\ref{eq:Nprop4}) in powers of $1/2m_N$ 
we arrive at  the expression 
given in appendix B of Ref.~\cite{hw07}\,:
\begin{eqnarray}
S_N (p\!-\!l)\!\!\!\!&=&\!\!\!
 \left(\frac{-i}{ v\!\cdot\!l}\right) 
\!\left[\frac{1\!+\!\gamma_0}{2} 
\!\left\{ \!1 \!-\;\!\frac{\vec{l}^{\; 2}\! -\!2\vec{l}\!\cdot\!\vec{p} }{2m_N \, l_0 }
\right\} 
\!- \;\! \!\frac{\vec{\gamma}\!\cdot\!(\vec{p} -\vec{l})}{2m_N}
\!+\frac{(v\!\cdot\!l)}{2m_N} \!\left(\!\frac{1\!-\!\gamma_0}{2}\right)\right] 
\!+\,\cdots\,\, 
\label{eq:Nprop5} 
\end{eqnarray} 
where the three terms containing the factor
$1/2m_N$ are of order $\tilde{\epsilon}$
relative to 1. 
In MCS, one applies the lowest-order 
non-relativistic approximation to eq.(\ref{eq:Nprop5})
(letting $\gamma_0 \to 1$ and ignoring the $\tilde{\epsilon}$
corrections)
and adopts as the heavy-nucleon propagator
\begin{eqnarray}
S_N (p\!-\!l)=-\,\frac{i}{v\cdot l} 
\label{eq:MCSprop} 
\end{eqnarray} 
In Ref.~\cite{hw07},
the $1/2m_N$ term within the curly brackets
in eq.~(\ref{eq:Nprop5}) 
was classified as a heavy-nucleon 
propagator recoil correction, while the other two
$1/2m_N$ terms were classified as higher order 
vertices in the the Lagrangian. 
In the present work, 
following Ref.~\cite{hw07}, we use 
eq.(\ref{eq:MCSprop}) without considering these  corrections.

\vspace{1mm} 
We investigate numerical differences in the 
transition amplitudes resulting from the use of 
the two different expressions,
eqs.(\ref{eq:Nprop1}) and (\ref{eq:MCSprop}),
for the nucleon propagator.
As mentioned,  the analytic expressions for the $pp$$\to$$pp\pi^0$
transition operators were calculated 
in Ref.~\cite{dkms99}
up to NNLO in the W-scheme
[i.e., the HB$\chi$PT propagator, eq.~(\ref{eq:Nprop1}) 
was used].\footnote{
The amplitude for diagram II given in 
Ref.~\cite{dkms99} has the wrong sign.}
It turns out that the amplitude expressions in MCS 
can be readily obtained from those given in Ref.~\cite{dkms99}.
The dependence on $v\!\cdot\!p$ in the 
amplitudes enters only through the HB$\chi$PT 
propagator eq.~(\ref{eq:Nprop1}) with $q=p - l$.  
This implies that the amplitudes resulting from the use of  
the MCS nucleon propagator, eq.~(\ref{eq:MCSprop}),  
can be simply 
obtained by formally replacing $v\!\cdot\!p$ with zero
($v\!\cdot\!p \!\to\!0$) in the expressions 
given in Ref.\cite{dkms99}. 
These expressions are valid for 
arbitrary kinematics but,  for the present purpose,
we simplify them with the use of the 
{\it fixed (frozen) kinematics approximation} (FKA),
wherein the external energies and momenta 
appearing in the particle propagators and vertices 
are ``frozen" at their threshold values.

\vspace{2mm}
In FKA, the operator corresponding to each
of the diagrams in Figs.~1$\sim\!$ 4 are written as
\begin{eqnarray}
T &=& \left( \frac{g_A}{f_\pi} \right)
\left( \vec{\Sigma}\cdot \vec{k} \right)
t(p,p^\prime , x)
\label{eq:T}
\end{eqnarray}
where $\vec{p}$ ($\vec{p}^{\; \prime}$)
is the relative three-momentum
in the initial (final) $pp$ state
($\vec{p}_1 - \vec{p}_2 = 2\vec{p}$,
$\vec{p}_1^{\; \prime} - \vec{p}_2^{\; \prime}
= 2\vec{p}^{\; \prime}$),
$\vec{k}\equiv\vec{p}-\vec{p}^{\; \prime}$,
$x=\hat{p}\cdot\hat{p}^\prime$, and
$\vec{\Sigma} = \frac{1}{2}
(\vec{\sigma}_1-\vec{\sigma}_2)$. 
The partial-wave projected form of $t(p,p',x)$ 
is written as~\cite{slmk}
\begin{eqnarray}
J[t]= -\left(\frac{m_Nm_\pi}{8\pi}\right)
\int_0^\infty \!p^2 {\rm d}p\; p^{\prime\; 2}
{\rm d}p^\prime
\int_{-1}^1 \!{\rm d}x\; \psi_{^1\!S_0}(p^\prime )\; 
t(p,p^\prime,x) 
 (p\!-\!p^\prime x) \psi_{^3\!P_0}(p)
\label{eq:J}
\end{eqnarray}
where $\psi_\alpha (p)$ is 
the $\alpha$ partial-wave
($^1\!S_0$ for the initial state and $^3\!P_0$
for the final state).\footnote{  
In this work we adopt
the plane-wave approximation, which corresponds to
the use of the wave functions of the generic form
$
\psi(p) =\delta (p- p_{on} ) / p^2  \; , 
$
where $p_{on}$ is the asymptotic on-shell nucleon 
momentum.} 
For the $t(p,p^\prime,x)$'s corresponding to the two-pion exchange (TPE)
diagrams (diagrams I, II, III and IV in Fig. 2),
it is informative to decompose each of them
into terms with definite {\it asymptotic} 
$k$-dependence as~\cite{ksmk07,sm99} 
\begin{eqnarray}
t(p,p^\prime , x) =\;
\;t_1\! \left( g_A/(8f_\pi^2)\right)^2| \vec{k} |
+ t_2\! \left(\ln\{|\vec{k} |^2/\Lambda^2\} \right)
+\, t_3 + \delta t(p,p^\prime , x),
\label{eq:tasympt}
\end{eqnarray}
where, in the limit of $k\!\to\!\infty$,
$t_3$ is $k$-independent,
and $\delta t(p,p^\prime , x) $ is  ${\cal O}(k^{-1})$.
The analytic expressions for $t_i$'s ($i=1, 2, 3$)
can be extracted~\cite{sm99} from
the amplitudes $T$ given in Ref.~\cite{dkms99}.
The first term with $t_1$ in eq.(\ref{eq:tasympt}) is
the ``leading part" of ${\cal O}(\tilde{\epsilon}^2)$ 
in MCS, which was evaluated in Ref.~\cite{hk02}; 
the remaining ``sub-leading" terms in eq.(\ref{eq:tasympt}) 
were not considered in Ref.~\cite{hk02}.
For the reason to be explained later,
we introduce the ``sub-leading" terms of 
${\cal O}(\tilde{\epsilon}^3)$ as 
\begin{eqnarray}
t^\star(p,p^\prime , x) \,\,\equiv\,\, t(p,p^\prime , x) 
-t_1\! \left( g_A/(8f_\pi^2)\right)^2| \vec{k} |
\label{eq:tstar}
\end{eqnarray}
and, correspondingly,
\begin{eqnarray}
J[t^\star]= -\left(\frac{m_Nm_\pi}{8\pi}\right)
\int_0^\infty \!p^2 {\rm d}p\; p^{\prime\; 2}
{\rm d}p^\prime
\int_{-1}^1 \!{\rm d}x\; \psi_{^1\!S_0}(p^\prime )\; 
t^\star(p,p^\prime,x) 
 (p\!-\!p^\prime x) \psi_{^3\!P_0}(p)
\label{eq:Jstar}
\end{eqnarray}

We remark that, for diagram I, we have
$t^\star(p,p^\prime,x) =t(p,p^\prime,x)$ 
and $J[t^\star]=J[t]$,
since the $t_1$ term exists
only for diagrams  II, III and IV~\cite{ksmk07,sm99}. 
In what follows, we investigate the numerical behavior 
of $J[t]$ and $J[t^\star]$.
For the sake of definiteness,
we concentrate here on a representative near-threshold case 
where the kinetic energy 
of the incident proton (in the laboratory system)
is $T_{lab} = 281$ MeV. 

We first consider the two-pion exchange (TPE) contributions,
coming from diagrams I $\sim$ IV 
depicted in Fig.~2.
Since the leading parts of the TPE diagrams of 
${\cal O}(\tilde{\epsilon}^2)$ are known 
to cancel among themselves~\cite{hk02,ksmk07},\footnote{
This can also be seen in our Table~\ref{one}.
Since $t_1 = 0 :-1:-1/2:3/2$ for diagram I, II, III and IV, respectively, 
the $J[t_1]$'s in the table
are found to be in the ratio of $0:-2:-1:3$ 
and  they add up to zero.
} 
our main concern here is the behavior 
of $t^\star(p,p^\prime , x)$, eq.(\ref{eq:tstar}),
or equivalently $J[t^\star]$, eq.(\ref{eq:Jstar}).
We show in Table~\ref{one} the values of $J[t^\star]$
corresponding to each of diagrams I $\sim$ IV. 
In the last column we 
also give $J[t]_{\rm Resc}$, which is the contribution
to $J[t]$ from the rescattering diagram (Resc) (Fig.1).
The second row gives 
$J[t^\star]$ calculated in MCS,
while the third row shows  
$J[t^\star]$ calculated in the W-scheme~\cite{ksmk07}.
Since the rescattering diagram contains no nucleon
propagators, the W-scheme and MCS give the same value
for $J[t]_{\rm Resc}$\,. 
The fourth row gives $J[t_1]$ representing
the contribution of the leading $t_1$ term; 
we remark that, for $J[t_1]$,  both MCS and the W-scheme 
give the same result. 
For comparison, we present in the fifth row
the value of $J[t]$ calculated in Ref.~\cite{dkms99}
with the use of the W-scheme.

Comparison of the $J[t^\star;MCS]$'s and 
$J[t^\star; W]$'s in Table~\ref{one}
indicates that the different treatments of the heavy-nucleon 
propagator between the W-scheme and MCS 
affect rather appreciably the individual values of $J[t^\star]$'s
for diagrams I $\sim$ IV.
Thus, for a formally consistent check of MCS
it is important to use the MCS nucleon propagator,
and in what follows we shall mostly concentrate on the results
obtained with the MCS nucleon propagators.

\begin{table} 
\caption{\label{one}\protect
The $pp\!\!\to\!\!pp\pi^0$ amplitudes $J$, 
eqs.~(\ref{eq:J}) and (\ref{eq:Jstar}),  
calculated for $T_{lab}= 281$ MeV
in the plane-wave approximation,
and in the frozen kinematics approximation (FKA).
The labels, I$\sim$IV, correspond to
the diagrams I$\sim$IV depicted in 
Fig.~2, while 
the column labeled ``Sum" gives
their combined contributions. 
The last column shows the contribution to $J[t]$
from the lowest-order non-vanishing one-pion-exchange 
rescattering diagram, the {\it Resc} diagram (Fig.1(a)).  
The second row shows $J[t^\star]$ 
calculated in MCS and formally of ${\cal O}(\tilde{\epsilon}^3)$,
while the third row gives $J[t^\star]$ evaluated 
in the W-scheme~\cite{dkms99}. 
The fourth row gives $J$ coming from the 
leading $t_1$ term of MCS order $\tilde{\epsilon}^2$;  
since $J[t_1;{\rm MCS}]
=J[t_1;{\rm W}]$ we  simply write $J[t_1]$.
The fifth row gives $J[t]$ calculated in Ref.~\cite{dkms99}
with the use of the W-scheme. 
$J[t^\star\!;{\rm W}]\!+\!J[t_1]\!=\!
J[t;{\rm  W} ]$, see eq.(\ref{eq:tstar}).
}
$$
\begin{array}{|l||r|r|r|r||c||c| }
\hline 
{\rm Type\, of \,diagram}  
&\!\! {\rm I}&{\rm II}\; \; &{\rm III}\; \; &{\rm IV} \; \; &
\;\:\:\;{\rm Sum} & {\rm Resc}\\
\hline \hline 
J[t^\star\!;{\rm MCS}]\!\times\!10^2 & -1.4 & -13.2 & -23.1 & 
12.5 & - 25.2 & 8\\
\hline
J[t^\star\!;{\rm W}]\!\times\!10^2
& -5.6   & -6.5 &  -29.9
& 4.9& - 37.1 &  8\\ 
\hline \hline 
J[t_1]\!\times\!10^2\! & 0 & -46.7& -23.3& 
70.0 & 0 &  \\
\hline 
J[t;{\rm  W} ]\!\times\!10^2 \; & -5.6 & -53.2  & -53.2 & 74.9  
& - 37.1  & 8\\ 
\hline 
\end{array}
$$
\end{table}

According to MCS, 
$J[t^\star]$ for the TPE diagrams should be of 
${\cal O}(\tilde{\epsilon}^3)$,
and $J[t]_{\rm Resc}$ 
should also be of ${\cal O}(\tilde{\epsilon}^3)$,
see Table~\ref{one}.
Thus, if MCS is a reasonable counting scheme,
$J[t^\star]$ and $J[t]_{\rm Resc}$
are expected to be of comparable magnitudes.
A comment is in order here, however, 
on the meaning of  ``comparable magnitudes".
Usually, two quantities that differ by a factor of
2$\sim$3 are considered to be of comparable magnitudes
(or of the same order) but,
in the present case where the expansion parameter is 
$\tilde{\epsilon}\sim 1/3$,
a difference by a factor of $\sim$3
may be interpreted as representing a different order in $\tilde{\epsilon}$.
Ideally speaking, one could make the situation clearer
by comparing terms that differ by two orders in $\tilde{\epsilon}$. 
In our present study, however, this is possible
only for certain classes of diagrams
(these cases will be discussed later in the text).
For the other cases in which MCS can be tested only within one order
in $\tilde{\epsilon}$, 
we shall adopt the following viewpoint:
If a deviation from the behavior expected from MCS 
lies within a factor of $\sim$3, we categorize it as  
{\it reasonably consistent} with MCS, 
in the sense that the deviation does {\it not} constitute
definitive evidence for a breakdown of MCS.

In looking at Table~\ref{one}, 
we first concentrate on diagrams II, III and IV,
leaving out diagram I for a while.  
We note that $J[t^\star;MCS]$ 
for diagrams II and IV 
are close to $J[t]_{\rm Resc}$, 
while $J[t^\star;MCS]$ for diagram III 
is larger than $J[t]_{\rm Resc}$ by a factor of $\sim$3.\footnote{
In what follows, we are mostly concerned with
the absolute value of $J$ rather than $J$ itself
but, for the sake of simplicity,
we shall refer to the absolute value of $J$ simply as $J$
(when there is no danger of confusion).} 
Thus the numerical behavior of the $J[t^\star;MCS]$'s
for diagrams II, III and IV 
is either consistent or reasonably consistent with MCS.
We find a similar situation in comparing  
$J[t^\star;MCS]$ (second row)
and $J[t_1]$ (fourth row).
According to MCS, 
$J[t^\star]$ should be one order higher in $\tilde{\epsilon}$
than $J[t_1]$.   
For diagrams II and IV,  $J[t^\star;MCS]$'s exhibit 
a slight over-suppression (beyond $\tilde{\epsilon}\sim$1/3)
but they are still {\it reasonably consistent} with MCS.
For diagram III,
$J[t^\star;MCS]\!\sim \!J[t_1]$ 
and hence no suppression is seen,
but the level of deviation from the MCS rule 
is again within a factor of $\sim$3,
presenting another {\it reasonably consistent} case.


According to Table~\ref{one},  $J[t^\star;MCS]$ for diagram I 
is smaller than $J[t]_{\rm Resc}$ by a factor of $\sim$6,
in contrast to the other TPE diagrams.
It is to be noted however 
that, for diagram I,  the leading term itself
exhibits an unusual behavior; {\it viz.,} 
although a straight-forward MCS counting 
indicates that diagram I, containing only
one nucleon propagator, should be of order 
$\tilde{\epsilon}^2$, the actual calculation
shows $t_1=0$~\cite{ksmk07}.\footnote{ 
This corresponds to the remark made in Ref.~\cite{hk02}
that diagram I is ``beyond next-to-leading 
order".}
It is thus likely that
some (possibly accidental) extra suppression mechanism
is at work for diagram I.
Since the existence of this type of extra suppression 
does not necessarily signal a breakdown of a perturbation series,
we may take the view that the results for diagram I 
is essentially consistent with MCS.
Based on the above discussion, 
we conclude that the overall behavior
of our numerical results shown in Table~\ref{one}
is {\it reasonably consistent} with MCS.
As mentioned, we are ignoring here 
the 
recoil correction to the 
MCS propagator eq.~(\ref{eq:MCSprop}). 
We will return to this issue later in the text.


\begin{table} 
\caption{\label{two}\protect
The $pp\!\!\to\!\!pp\pi^0$ amplitudes $J[t]$, 
eq.~(\ref{eq:J}),  
calculated for $T_{lab}= 281$ MeV
in the plane-wave approximation and in FKA.
The labels, V, VI and VII, correspond to
the diagrams V, VI and VII depicted in 
Figs. 3 and 1, respectively.
The last column gives
the contribution to $J[t]$ from of the rescattering diagram (Fig.1). 
The second and third rows show,
respectively, $J[t]$ calculated in MCS and in the W-scheme~\cite{dkms99}. 
}
$$
\begin{array}{| l || c | c | c|| c |}
\hline 
{\rm Type\, of \,diagram}  
 & {\rm V} & {\rm VI} & {\rm VII} & {\rm Resc}\\
\hline \hline 
J[t;{\rm MCS}]\!\times\!10^2 &\; 3.4 \;& \;-2.3\; &\; 20.8\; & 8\\
\hline
J[t;{\rm W}]\!\times\!10^2
&\; 1.4\; & \;1.1\; & \;20.8\;  &  8\\ 
\hline
\end{array}
$$
\end{table}

We next discuss the behavior of 
the other higher-order contributions
coming from diagrams V and VI (Fig.\ 3) and diagram VII (Figs.\ 1(b)).
In diagram VII the 
pion-nucleon rescattering vertex is given by 
${\cal L}_{\pi N}^{(3)}$, see eq.(C.3) 
in Ref.~\cite{fms98};
this vertex contains the recoil correction 
to the pion-nucleon vertex in the lower 
chiral order re-scattering amplitude ({\it Resc}) in Fig.\ 1(a). 
In Table~\ref{two} we show $J[t]$ for diagrams V, VI and VII,
calculated in MCS (second row) and 
in the W-scheme (third row);
this latter has been taken from Ref.\cite{dkms99}.
According to MCS~\cite{hanhart04}, 
diagrams V and VI, which are 
the pion s-wave re-scattering diagrams containing 
a pion loop at one nucleon vertex,  
should be suppressed by 
$\tilde{\epsilon}^2 \simeq 1/7$ compared to diagram VII.
The numerical results in Table~\ref{two} 
are in conformity with
this expected suppression. 
Meanwhile, diagrams VII and Resc (shown in Fig.1)
should both be of order $(\tilde{\epsilon})^3$
in MCS~\cite{hk02,hanhart04}.
Table~\ref{two} indicates that 
$J[t]$ for diagram VII is larger than $J[t]_{\rm Resc}$ 
by a factor of $\sim$3, 
but here again this level of deviation from the MCS prediction 
is regarded as constituting a {\it reasonably consistent} case.
It is informative to study in more detail the origin 
of the differences between these two contributions.
For threshold kinematics (FKA), the pion-nucleon 
re-scattering vertices in the {\it Resc} diagram and diagram VII 
are given, respectively, by\footnote{
For the vertex for the {\it Resc} diagram, see eq.(A.29) in 
Ref.~\cite{bkm95}; for the vertex for diagram~VII,
see eq.(C.3) in Ref.~\cite{fms98}. } 
\begin{eqnarray}
{\rm Resc:} 
\;\;\;\;\;\;\;\;\;\;\;\;\;\;&& -i\frac{m_\pi^2}{f_\pi^2}
\left[ 4c_1- \left(c_2-\frac{g_A^2}{8m_N}\right) -c_3\right] 
\label{eq:Resc}\\
{\rm VII:} \;\;\;\;\;\;\;\;\;\;\;\;\;\;&& 
i\frac{m_\pi^2}{f_\pi^2}\left[c_2\left(1+
\frac{3m_\pi}{4m_N}\right) -\frac{g_A^2}{16m_N}\right] 
\label{eq:seven}
\end{eqnarray}
Formally, these two vertices do seem to be of the same order in MCS.
However, the well-known smallness of the $\pi N$ iso-scalar 
scattering length requires  
a substantial cancellation between the three 
$c_i$ terms in eq.(\ref{eq:Resc}).
This particular situation causes the expression 
in eq.(\ref{eq:seven}) to become 
larger than that in eq.(\ref{eq:Resc})
by almost a factor of  3. 
These features illustrate intrinsic subtlety we encounter
in discussing the convergence property of MCS
beyond the level of a factor of $\sim$3.   
On the other hand, according to the W-scheme,
the {\it Resc} diagram belongs to NLO
and diagram VII  to NNLO (see Table 1).
Our numerical results exhibit
a definite deviation from the prediction based on the W-scheme.

We have seen above that, despite the fact
that the expansion parameter $\tilde{\epsilon}$
in MCS is not very small ($\tilde{\epsilon}\sim 1/3$), 
MCS is likely to provide a useful guide
in organizing the higher-order terms.
In this connection, it seems worthwhile
to examine $J[t]$ as a function of $\tilde{\epsilon}$
by allowing $\tilde{\epsilon}$ to vary from its physical value
to smaller values.
This may offer information somewhat different from
that obtainable by comparing the magnitudes of the individual diagrams
at the fixed value of $\tilde{\epsilon}$.
We explore here how the contributions of the diagrams 
behave when the pion mass $m_\pi$ is artificially varied. 
To extract the non-trivial $m_\pi$-dependence of $J$, 
we define the plane-wave amplitude $X$ by
\begin{eqnarray}
J[t^\star]&=& -\left(\frac{m_Nm_\pi}{8\pi}\right)
2 | \vec{k} | \,X 
\label{eq:X}
\end{eqnarray} 
As for the ``leading term" of the TPE amplitude 
(viz., the $t_1$-term contribution), 
we take as a representative the $J[t_1]$ for diagram II
and introduce
\begin{eqnarray}
J[t_1;{\rm diagram\,II}]
&\equiv& -\left(\frac{m_Nm_\pi}{8\pi}\right) 
2 | \vec{k} | \, X_{t_1} 
\label{eq:Xt1a}
\end{eqnarray} 
which gives ($t_1= 1$ for diagram II) 
\begin{eqnarray}
X_{t_1} &=& 
\left( g_A/(8f_\pi^2)\right)^2| \vec{k} | \; .
\label{eq:Xt1b} 
\end{eqnarray}
In Table~\ref{three} we show as functions of $m_\pi$
the values of $X$ for individual diagrams calculated in MCS.
The graphical representation of $X/m_\pi$ as a function of $m_\pi$
is given in Fig.~5.
We expect from eq.(\ref{eq:Xt1b}) that 
$X_{t_1} \simeq \sqrt{m_\pi}$ for $m_\pi\!\to\! 0$,  
and the numbers in 
the second column of Table~\ref{three} 
indicate that this expectation is borne out.
According to MCS, the $X$'s for diagrams II, VII 
and the {\it Resc} diagram are expected 
to be linear in $m_\pi$ for $m_\pi\!\to\! 0$.
The numerical results in Table~\ref{three} 
are in agreement with this MCS expectation. 
According to Fig.~5,
the $X$s for the TPE diagrams I, III and IV appear 
to have a more complex $m_\pi$ dependence than 
implied by a naive application of MCS.
The $X$s here diminish less rapidly for $m_\pi\!\to \!0$ than 
the expected linear dependence on $m_\pi$.
We remark that in MCS 
the chiral log factor, ln$(m_\pi )$, 
which arises from the loops of the TPE diagrams,
is treated as a constant factor that does not affect 
the MCS power counting.
Meanwhile, the existence of the $t_2$-term in eq.(\ref{eq:tasympt})
implies that the $m_\pi$-dependence in the chiral log
can disturb the simple power counting in MCS, 
and Fig.~5 indicates the chiral log behavior is dominant at 
low $m_\pi$ values.
%
Finally, according to MCS, the $X$s for diagrams V and VI 
are expected to vary as $m_\pi^2$ for $m_\pi\!\to\! 0$.
Our numerical results indicate 
that they go to zero faster than $m_\pi$ but 
somewhat slower 
than $m_\pi^2$. 

\begin{table} 
\caption{\label{three}\protect 
The plane-wave amplitude $X$ [eq.(\ref{eq:X})]
(in units of $10^{-2}$fm$^3$) 
calculated for $T_{lab}= 281$ MeV in FKA;
e.g., the second column gives the values of $X_{t_1}$, eq.(\ref{eq:Xt1b}). 
For the cases where MCS and the W-scheme give different results,
MCS is used.}
$$
\begin{array}{|c||r| |r|r|r|r|r| |r|r|r| }
\hline
{\rm Diagram }\: & 
t_1\;& t^\star{\rm :I} &
\,t^\star{\rm :II} &\;t^\star{\rm :III}& t^\star{\rm :IV} &
t{\rm :Resc} &t{\rm :V}&t{\rm :VI} & t{\rm :VII} \\
\hline \hline 
  m_\pi \!=\!140\,{\rm MeV}  & 92.6 & 2.7& 26.2  & 45.9 & -24.8  
&-15.7&-6.8 & 4.6 & -41.5\\ 
\hline 
100 & 77.9 &1.5 & 18.6 & 43.8& -19.5
& -11.6 & -3.8 & 2.6 &-29.9\\
\hline 
50  &54.7 & 0.3 & 9.4&  32.6 & -11.3
 & -6.1 & -1.1  & 0.8 &-15.1 \\ 
\hline 
10 & 24.3 &-0.1& 2.1 &10.9 & -2.9 
& -1.3 & -0.06 & 0.04  & -3.1\\
\hline
\end{array}
$$
\end{table}

\begin{figure}
\begin{center}
\includegraphics[width=10cm]{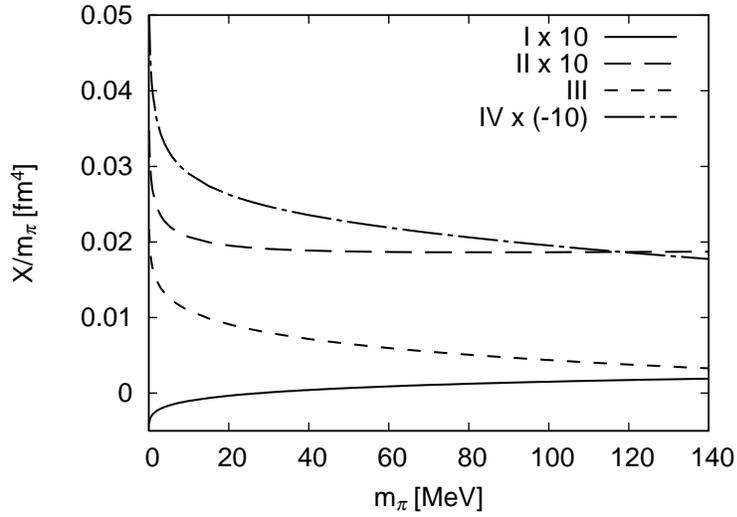}
\caption[]{\protect
For each of the TPE diagrams, $X/m_\pi$ is plotted
as a function of $m_\pi$;
the amplitudes $X$ are defined in eq.(\ref{eq:X}). 
}
\end{center}
\end{figure}

Besides the diagrams so far considered,
the five-point vertex counter-term diagram 
(CT diagram)
shown in Fig.~4 is also of order 
$\tilde{\epsilon}^3$ in MCS~\cite{hanhart04},
and we now give a brief discussion of the numerical behavior 
of the CT diagram.
An aspect that distinguishes this diagram from the others 
is that the LEC here is an unknown parameter whereas
the LECs in the other diagrams are predetermined from 
separate sources of information.
In the absence of experimental data
needed to determine the relevant LEC, we rely here 
on the resonance saturation prescription 
and use the $\sigma$- and $\omega$-exchange mechanism 
considered in Lee and Riska's work~\cite{lr93}. 
In this model, the amplitude $T$ [eq.(\ref{eq:T})]
for the CT diagram is given by, see e.g. Ref.~\cite{cfmv96}
\begin{eqnarray}
T_{CT} &=& \frac{ g_A \omega_q }{2f_\pi m_N^2} 
\left[ \left( \frac{ g_\sigma^2 }
{ m_\sigma^2 } + 
\frac{ g_\omega^2 }{ m_\omega^2 } 
\right)
\vec{\Sigma}\cdot \vec{P} 
\right. 
\nonumber \\ && 
\left. 
\;\;\;\;\;\;\;\;\;\;\;\;\;\;\;\;\;\;\;\;
-\, i\, \frac{ g_\omega^2 }{ m_\omega^2 }
\left( 1+C_\omega \right) 
\left(\vec{\sigma}_1 \!\times \!\vec{\sigma}_2 \right)\!\cdot\! \vec{k}\;
\right] 
\label{eq:CT}
\end{eqnarray}
where $\omega_q =\sqrt{m_\pi^2 + \vec{q}^{\; 2} }$ 
is the energy of the outgoing pion and 
$\vec{P}=\vec{p}+\vec{p}^{\; \prime}$.\footnote{
Between the spin singlet and triplet states
(between $^1S_0$ and $^3P_0$),
the operator 
$i\left(\vec{\sigma}_1 \times \vec{\sigma}_2 \right)$
is equivalent to $(-2) \vec{\Sigma}$.}
A caveat here is that, since the ``$\sigma$" exchange 
is considered to represent a scalar part of 
correlated two-pion-exchange in the $NN$ potential,
and since we have already taken into account some TPE diagrams, 
there is a danger of double-counting
if we include the entire ``$\sigma$" exchange
contribution to the CT diagram.
Without going into this issue, we present here
the individual contributions of the ``$\sigma$"- 
and $\omega$-exchange 
and compare them with the contributions 
of the other diagrams that are expected to be 
of order $\tilde{\epsilon}^3$ in MCS.
If again we restrict ourselves to threshold kinematics
($\vec{p}^{\; \prime} =0$ and $\vec{k}=\vec{P}=\vec{p}$), 
eq.(\ref{eq:CT}) can be effectively rewritten as
\begin{eqnarray}
T_{CT} = \left( \frac{ g_A  }{f_\pi } \right) 
\vec{\Sigma} \cdot \vec{p} \left(\frac{m_\pi}{2m_N^2}\right)
\left\{ \frac{ g_\sigma^2 } { m_\sigma^2 } +
\frac{ g_\omega^2 }{ m_\omega^2 } 
\left[ 1 \!+ \!2 \left( 1\!+\!C_\omega \right) \right] 
\right\} 
\label{eq:CT1}
\end{eqnarray}
which, with the use of 
$g_\sigma^2/(4 \pi)\!= \!5$,
$g_\omega^2/(4 \pi)\!=\!10$, 
$m_\sigma \!=\!600$ MeV and 
$m_\omega\!=\!780$ MeV, leads to
\begin{eqnarray}
T_{CT} = \left( \frac{ g_A  }{f_\pi } \right) 
\vec{\Sigma} \cdot \vec{p} 
\left\{13.8 \,{\rm GeV}^{-3} +
16.4\,{\rm GeV}^{-3} \left[ 1  
\!+ \!2 \left( 1\!+\!C_\omega \right) \right] 
\right\} 
\label{eq:CT2}
\end{eqnarray} 
Referring to the relation between $T$ and $J$, 
see eqs.~(\ref{eq:T}) and (\ref{eq:J}),
we estimate $J_{CT}$ from eq.(\ref{eq:CT2})  to be 
\begin{eqnarray}
J_{CT} &=& - 0.052 - 0.062\,[3+2C_\omega ] 
\end{eqnarray} 
where the first (second) term comes from
$\sigma$-exchange ($\omega$-exchange).\footnote{
In the literature we find $| C_\omega | < 1$ but its value is 
not well determined;  see, e.g., Refs.~\cite{pm02, sl96}. 
} 
The magnitude of the $J_{CT}$ amplitude is to be compared to 
$J(t^\star ; MCS)$ in 
the second row in Table~\ref{one}.  
Within the factor of $\sim 3$ the 
$\sigma$- and $\omega$-exchange contributions 
are comparable to the 
${\cal O}(\tilde{\epsilon}^3)$ TPE amplitudes in Table 2. 
Although, as stated earlier, 
there is a question of possible double-counting 
concerning the ``$\sigma$"-exchange contribution,
it is to be noted that the CT amplitude coming from 
$\omega$-exchange alone is of expected MCS magnitude.

\vspace{3mm}
To summarize, we have calculated contributions to 
the threshold $pp\to pp\pi^0$ reaction amplitude
with the use of the MCS heavy-baryon propagator, 
studied the convergence property of the
expansion based on MCS, 
and compared the results with those obtained in the W-scheme. 
We have considered all the diagrams
up to NNLO in the W-scheme
and, taking advantage of the fact that 
these diagrams belong to widely different orders in MCS
(ranging ${\cal O}(\tilde{\epsilon}^2)$ 
to ${\cal O}(\tilde{\epsilon}^5)$),
we have examined whether the actual numerical behavior
of these diagrams substantiates the MCS predictions.
Our numerical results indicate:
(a) The terms that are expected to be of the same order
in MCS follow that pattern within a factor of $\sim3$;
(b) The terms that are expected to differ by one order 
in $\tilde{\epsilon}$ in MCS also follow that pattern
within a factor of $\sim3$;
(c) The terms that are predicted by MCS to be suppressed
by two orders in $\tilde{\epsilon}$ 
($\tilde{\epsilon}^2\simeq1/7)$
clearly show the expected suppression.
Thus all the diagrams studied here exhibit numerical behaviors
that are (in our definition) {\it reasonably consistent} 
[Cases (a) and (b) above],
or consistent [Case (c)].
These results lead us to conclude the following
(i) Even though the expansion parameter $\tilde{\epsilon}$
in MCS is not very small ($\tilde{\epsilon}\sim 1/3$), 
MCS provides a useful semi-quantitative guide
in organizing the higher-order terms;
(ii) MCS is superior to the W-scheme
in organizing the relative importance of 
the higher order terms. 
This last conclusion supplements the finding in Ref.~\cite{hvm00}, 
which shows that MCS converges for p-wave pion production. 
%
%

A better test of the convergence property 
of MCS would be to calculate next-order contributions 
for all the diagram under consideration
and compare the terms that are expected to differ by two orders 
of $\tilde{\epsilon}$, but this is beyond the scope
of the present work.
As mentioned we have neglected the 
${\cal O}(\tilde{\epsilon}^3)$ contributions 
to the reaction amplitude
arising from  the recoil correction to the MCS propagator,
eq.(\ref{eq:MCSprop}), and we have also ignored the additional vertices
discussed in connection eq.(\ref{eq:MCSprop}).
We remark, however, that these contributions 
do not directly affect the behavior of the contributions
of the diagrams studied in the present investigation. 
%

\vspace{3mm}

Correspondence with C. Hanhart is gratefully acknowledged. 
This work is supported in part
by the US National Science Foundation,
Grant Nos.  PHY-0457014 and PHY-0758114, and
by the Japan Society for the Promotion of Science,
Grant-in-Aid for Scientific Research (C) No.20540270.


\begin{thebibliography}{99}

\bibitem{meyeretal90}
H.O. Meyer {\it et al.},
Phys. Rev. Lett. {\bf 65}, 2846 (1990);
Nucl. Phys. A, {\bf 539}, 633 (1992).

\bibitem{ms91}
G.A. Miller and P.U. Sauer,
Phys. Rev. C, {\bf 44}, R1725 (1991).

\bibitem{lr93}
T.-S.H. Lee and D.O. Riska,
Phys. Rev. Lett. {\bf 70}, 2237 (1993);
see also C.J. Horowitz, H.
O. Meyer and D.K. Griegel, Phys. Rev. C, {\bf 49}, 1337 (1994).

\bibitem{pmmmk96}
B.-Y. Park, F. Myhrer, J.R. Morones, T. Meissner
and K. Kubodera, Phys. Rev. C, {\bf 53}, 1519  (1996).

\bibitem{cfmv96}
T.D. Cohen, J.L. Friar, G.A. Miller and U. van Kolck,
Phys. Rev. C, {\bf 53}, 2661 (1996).

\bibitem{am01}
S. Ando, T.S. Park and D.P. Min,
Phys. Lett. B, {\bf 509}, 253 (2001).

\bibitem{hk02}
C. Hanhart and N. Kaiser, Phys. Rev. C,
{\bf 66}, 054005 (2002). 

\bibitem{hanhart04}
C. Hanhart, Phys. Rep. {\bf 397}, 155 (2004)

\bibitem{Lensky05}
V. Lensky, J. Haidenbauer, C. Hanhart,
V. Baru, A. Kudryavtsev and U.-G. Meissner,
Eur. Phys. J. A, {\bf 27}, 37 (2006) [nucl-th/0511054];
V. Lensky {\it et al.}, Int.J.Mod.Phys. {\bf A22}, 591 (2007) 
[nucl-th/0609007]. 

\bibitem{fm06}
F. Myhrer,
``Large two-pion-exchange contributions to the
$pp\to pp\pi^0$ reaction", in Conf. Proc.
{\it Chiral Dynamics 2006}, 
eds. M.W. Ahmed, H. Gao, B. Holstein and H.R. Weller
(World Scientific, Singapore, 2007),
[arXiv:nucl-th/0611051].

\bibitem{ksmk07}
Y. Kim, T. Sato, F. Myhrer and K. Kubodera,
Phys. Lett. {\bf B 657}, 187 (2007).

\bibitem{Baruetal08} 
V. Baru {\it et al.}, Phys. Lett. B {\bf 659}, 184 (2008).

\bibitem{slmk}
T. Sato, T.-S.H. Lee, F. Myhrer and K. Kubodera,
Phys. Rev. C, {\bf 56}, 1246 (1997). 

\bibitem{hvm00}
C. Hanhart, U. van Kolck and G.A. Miller, 
Phys. Rev. Lett. {\bf 85}, 2905 (2000)

\bibitem{dkms99}
V. Dmitra\v{s}inovi\'{c}, K. Kubodera, F. Myhrer and T. Sato,
Phys. Lett. B, {\bf 465}, 43 (1999).

\bibitem{bkm95} 
V. Bernard, N. Kaiser and U.-G. Meissner, 
Int. J. Mod. Phys. {\bf E 4}, 193 (1995).

\bibitem{weinberg91}
S. Weinberg, Nucl. Phys. B {\bf 363}, 3 (1991)

\bibitem{hw07}
C. Hanhart and A.Wirzba, 
Phys. Lett., {\bf B 650}, 354 (2007) 
[arXiv:nucl-th/0703012].

\bibitem{sm99}
T. Sato and F. Myhrer,
unpublished notes (1999).

\bibitem{fms98} 
N. Fettes, U.-G. Meissner and S. Steininger,
Nucl. Phys. {\bf A 640}, 199 (1998) 
[arXiv:hep-ph/9803266]

\bibitem{pm02} G. Penner and U. Mosel, Phys. Rev. C {\bf 66}, 055211 (2002). 

\bibitem{sl96} 
T. Sato and T.-S. H. Lee, Phys. Rev. C {\bf 54}, 2660 (1996).




\end{thebibliography}
\end{document}